# High-pressure study of the non-Fermi liquid material $U_2Pt_2In$

P. Estrela[1], A. de Visser[1,*], T. Naka[1,2], F.R. de Boer[1] and L.C.J. Pereira[3]

[1] *Van der Waals-Zeeman Institute, University of Amsterdam,*
*Valckenierstraat 65, 1018 XE Amsterdam, The Netherlands*
[2] *National Research Institute for Metals, 1-2-1 Sengen, Tsukuba, Ibaraki 305-0047, Japan*
[3] *Department of Chemistry, Technological and Nuclear Institute,*
*Apartado 21, 2686-953 Sacavém, Portugal*

The effect of hydrostatic pressure ($p \leq 1.8$ GPa) on the non-Fermi liquid state of $U_2Pt_2In$ is investigated by electrical resistivity measurements in the temperature interval 0.3-300 K. The experiments were carried out on single-crystals with the current along ($I \parallel c$) and perpendicular ($I \parallel a$) to the tetragonal axis. The pressure effect is strongly current-direction dependent. For $I \parallel a$ we observe a rapid recovery of the Fermi-liquid $T^{\,2}$-term with pressure. The low-temperature resistivity can be analysed satisfactorily within the magnetotransport theory of Rosch, which provides strong evidence for the location of $U_2Pt_2In$ at an antiferromagnetic quantum critical point. For $I \parallel c$ the resistivity increases under pressure, indicating the enhancement of an additional scattering mechanism. In addition, we have measured the pressure dependence of the antiferromagnetic ordering temperature ($T_N = 37.6$ K) of the related compound $U_2Pd_2In$. A simple Doniach-type diagram for $U_2Pt_2In$ and $U_2Pd_2In$ under pressure is presented.

* Corresponding author:    Dr. A. de Visser
Van der Waals-Zeeman Institute, University of Amsterdam
Valckenierstraat 65, 1018 XE Amsterdam, The Netherlands
Phone: +31-20-5255732, Fax: +31-20-5255788
E-mail: devisser@wins.uva.nl

# I. INTRODUCTION

$U_2T_2X$ intermetallics, where T is a transition metal and X is In or Sn, have been the subject of intensive research, as this family of compounds may serve as an exemplary series to study the systematics of 5$f$-electron hybridization [1]. The hybridization strength can be tuned by choosing the appropriate T and X elements, and as a result various ground states are observed, e.g. Pauli paramagnetism, local-moment antiferromagnetism and pronounced spin-fluctuating behavior. Among the $U_2T_2X$ compounds, $U_2Pt_2In$ takes a special place, because: (i) it is a non-ordering heavy-electron compound with a strongly renormalized quasiparticle mass ($c/T = 0.41$ J/molU-K$^2$ at $T = 1$ K) [1] and (ii) it shows pronounced departures from the standard Fermi-liquid (FL) behavior, or, in other words, it is a non-Fermi liquid (NFL) compound [2,3]. Currently, NFL compounds attract much attention [4-6], because NFL behavior may be considered to represent a new ground state. In the case of $U_2Pt_2In$, the NFL properties are summarized by: (i) the specific heat varies as $c(T) \sim -T\ln(T/T_0)$ over almost two decades of temperature ($T = 0.1$-6 K) [2], (ii) the magnetic susceptibility shows a weak maximum at $T_m = 8$ K for a magnetic field along the $c$ axis (tetragonal structure), while it increases as $T^{0.7}$ when $T \to 0$ for a field along the $a$ axis [3], and (iii) the electrical resistivity obeys a power law $T^\alpha$ with $a = 1.25 \pm 0.05$ ($T < 1$ K) and $0.9 \pm 0.1$ ($T \to 0$), for the current along the $a$ and $c$ axis, respectively [7]. It is important to realize that $U_2Pt_2In$ is one of the rare *stoichiometric* (undoped) compounds which exhibits NFL behavior *at ambient pressure*. This has the advantage that NFL properties can be examined without the need to apply mechanical pressure, like in $CePd_2Si_2$ and $CeIn_3$ [8], or chemical pressure, like in $Ce(Cu,Au)_6$ [6]. Other stoichiometric NFL compounds are $CeNi_2Ge_2$ [9], $CeCu_2Si_2$ [10], $YbRh_2Si_2$ [11] and $U_3Ni_3Sn_4$ [12].

The origin of the NFL behavior in $U_2Pt_2In$ is still not settled, despite a thorough experimental characterization [7]. The most plausible scenarios are: (i) the proximity to a magnetic quantum critical point (QCP) and (ii) Kondo disorder. The QCP scenario is often discussed in terms of the Doniach-type of phase diagram [13], which describes the competition between magnetic order and Kondo screening. Magnetic order emerges when the inter-site Ruderman-Kittel-Kasuya-Yosida (RKKY) interaction energy, given by k$_B T_{RKKY}$, starts to dominate the single-ion Kondo interaction energy, k$_B T_K$. By controlling the ratio $T_{RKKY}/T_K$ by varying the strength of the $f$-electron hybridization, the compound might be tuned to a magnetic QCP at $T = 0$. The QCP controls the physics over a wide range of temperatures, which results in NFL behavior. Expressions for the low-temperature NFL term in the thermal, magnetic and transport properties of an itinerant (anti)ferromagnet, tuned to its quantum critical point, have been evaluated by Millis [14] and Rosch [15]. These expressions are equivalent to the ones obtained by the self-consistent renormalization theory of spin fluctuations in itinerant electron systems derived by Moriya and Takimoto [16]. In the case of $U_2Pt_2In$, the QCP scenario is supported by the notion that the compound is located at the non-magnetic side, close to the magnetic/non-magnetic borderline, in the Doniach-type phase diagram for the $U_2T_2X$ family of compounds [7,17]. The absence of magnetic order in $U_2Pt_2In$, at least down to $T = 0.05$ K, was recently demonstrated by μSR experiments [18], which put an upper bound of ~ 0.1 Oe on the internal field due to weak magnetic order.

The second possible explanation for NFL behavior in $U_2Pt_2In$, is the presence of Kondo disorder [19]. Large disorder in a material may produce a distribution of Kondo temperatures. For each single-magnetic impurity the Kondo effect will take place at a different value of $T_K$. Averaging over such a distribution may result in thermodynamic and transport properties with NFL-like dependencies, due to the broad range of effective Fermi temperatures. Indeed, the residual resistivity values of $U_2Pt_2In$ are substantial [3], i.e. of the order of 100 μΩcm, which normally indicates significant crystallographic disorder. However, Rietveld analyses of the diffraction patterns obtained by single-crystal x-ray [3] and neutron-diffraction [20], do not confirm the presence of significant disorder. In fact, the refinement factors are good and exclude significant Pt-In site inversion.

Hydrostatic pressure is a convenient tool to probe the location of compounds in the Doniach-type diagram. In general, pressure reduces the ratio $T_{RKKY}/T_K$, which results in the suppression of magnetic order. In the case of $U_2Pt_2In$, pressure is expected to drive the material further into the non-magnetic regime, away from the QCP. This should result in the recovery of the Fermi-liquid state.

Recently, Rosch [15] has presented a theory of magnetotransport in correlated metals near an antiferromagnetic QCP, which delineates the NFL and Fermi-liquid regimes as a function of the distance to the QCP and the amount of disorder in the material. The theory is based on the assumption - for heavy-fermion systems - that the low-energy excitations below a characteristic temperature $T_K$ are heavy quasiparticles and their excitations. In the vicinity of the QCP, the resistivity is then determined by scattering of quasiparticles by spin fluctuations.

These scattering processes are most important near hot lines, i.e. points on the Fermi surface connected by **Q**, where **Q** is the magnetic ordering vector of the antiferromagnetic phase. Considering that spin fluctuations are destroyed at the temperature scale *G*, where *G* is typically of the order of $T_K$ or the coherence temperature $T_{coh}$, a universal resistivity is derived. Using the dimensionless parameters $t = T/G$ as a measure for the temperature, $x = r_0/r_m \approx 1/RRR$ (Residual Resistance Ratio) as a measure for the amount of disorder and $r \propto (d - d_c)/d_c$ as a measure for the distance to the QCP in the paramagnetic phase, the resistivity is universal for $t < x^{1/2}$ and $r < 1$ in the scaling limit $t, x, r \to 0$ and $t/x, r/x \to$ const. Here $r_0$ is the residual resistivity, $r_m$ is a typical high-temperature ($t \sim 1$) resistivity value and *d* is a control parameter with critical value $d_c$. Three different regimes are predicted for the resistivity $\Delta r = r - r_0$. For the three dimensional case ($d = 3$) these are: (I) $\Delta r \sim t^{3/2}$, (II) $\Delta r \sim t x^{1/2}$ and (III) $\Delta r \sim t^2 r^{-1/2}$. The temperature ranges of different regimes depend on the amount of disorder in the system. For very clean samples ($x \ll 1$) the $\Delta r \sim t^{3/2}$ dependence is observed at very low temperatures. In the dirty limit ($x \to 1$) region II is not observed, i.e. no $r \sim T$ regime occurs, while regimes I and III, which are called the disorder-dominated regime and the disorder-dominated Fermi liquid regime, respectively, extend over a large temperature range.

In this paper we report a high-pressure transport study on single-crystalline $U_2Pt_2In$. The electrical resistivity, $r(T)$, was measured for a current, *I*, along the *a* and *c* axis, up to pressures of 1.8 GPa. Here we concentrate on the results obtained for $I \parallel a$. The results for $I \parallel c$ have partly been reported in Ref.[21]. We show, for $I \parallel a$, that, while $r \sim T^\alpha$ with *a* = 1.25±0.05 for $T < 1$ K at ambient pressure, the Fermi-liquid $T^2$ term is rapidly recovered under pressure. The analysis of the pressure dependence of the FL temperature interval within the theory of Rosch is consistent with $U_2Pt_2In$ being situated at or very close to an antiferromagnetic quantum critical point. In order to obtain an estimate of the strength of the control parameter in the Doniach-type phase diagram of the $U_2T_2X$ compounds, we have also measured the pressure dependence of the antiferromagnetic transition temperature $T_N$ = 37.6±0.5 K of the heavy-fermion material $U_2Pd_2In$. In this compound the uranium moments are confined to the basal plane and form a non-collinear magnetic structure [22]. The electronic specific-heat coefficient *g* amounts to 0.20 J/molU-$K^2$ [1]. Under pressure, $T_N$ decreases slightly to 35.2±0.5 K at 1.8 GPa, which is consistent with the simple Doniach-diagram picture.

## II. EXPERIMENTAL

A single-crystalline batch of $U_2Pt_2In$ was prepared by a modified mineralization technique [23]. $U_2Pt_2In$ is a polymorphic compound, as was recently shown in Ref.[3]. Polycrystalline samples crystallize in the tetragonal $U_3Si_2$-type of structure (space group P4/*mbm*) with lattice parameters $a = 7.654$ Å and $c = 3.725$ Å, while single-crystalline material forms in the tetragonal $Zr_3Al_2$-type of structure (space group P4$_2$/*mnm*) with lattice parameters $a = 7.695$ Å and $c = 7.368$ Å. The $Zr_3Al_2$-type structure can be considered as a super-structure of the $U_3Si_2$-type of structure, with a doubling of the c-axis. Despite this polymorphism, significant differences in the electronic and magnetic properties of single- and polycrystalline samples have not been observed [7]. The residual resistivity values are large: $r_0$ equals 115 µΩcm and 210 µΩcm, for $I \parallel a$ and $I \parallel c$, respectively [3]. Since the resistivity at room temperature amounts to 220 µΩcm, low residual resistance ratio's $r_{RT}/r_0$ (where $r_{RT} \equiv r(300K)$) result: 1.9 and 1.1 for the *a* and *c* axis, respectively. The relative errors in these numbers amount to 10%, because of the uncertainty in the determination of the geometrical factor in the resistivity experiment. High $r_0$-values normally indicate significant crystallographic disorder. However, as mentioned in the previous section, single-crystal x-ray [3] and neutron-diffraction [20] experiments do not confirm the presence of significant disorder. Transport experiments in a magnetic field lead to a reduction of $r_0$, which indicates that at least part of the high $r_0$-value is intrinsic and not due to defects and/or impurities [7]. The single-crystalline sample of $U_2Pd_2In$ was also prepared by the mineralization method. $U_2Pd_2In$ crystallizes in the tetragonal $U_3Si_2$-type of structure with lattice parameters $a = 7.637$ Å and $c = 3.752$ Å.

The electrical resistivity of $U_2Pt_2In$ under pressure ($p \leq 1.8$ GPa) was measured for $I \parallel a$ and $I \parallel c$ in the temperature range 0.3-300 K. The resistivity was measured on bar-shaped or platelet-like samples using a standard low-frequency four-probe ac-technique with a typical excitation current of ~ 100 µA. The resistivity under pressure was measured using a copper-beryllium clamp cell. The samples were mounted on a specially designed plug and inserted into a teflon holder together with the pressure transmitting medium. A short tungsten carbide piston is used to transfer the pressure to the teflon holder. A mixture of Fluorinerts was used as pressure transmitting medium. The pressure values (accuracy 0.05 GPa) were calculated from the external load and corrected for an empirically determined efficiency of 80%. The pressure dependence of $r_{RT}$ was negligible. However, small changes in the geometrical factor (mainly in the distance between the voltage contacts) sometimes occurred. Therefore, at each pressure, the resistance curves were normalized to 1 at room temperature. The electrical resistivity of $U_2Pd_2In$ under pressure was measured in a similar way for a current along the [101] direction.

In addition, the isothermal compressibility $k = -V^{-1}(dV/dp)_T$ of $U_2Pt_2In$ was measured on a powdered single-crystalline sample by x-ray diffraction under pressure up to 7 GPa.

### III. RESULTS

The electrical resistivity $r_a(T)$ and $r_c(T)$ of $U_2Pt_2In$, normalized to 1 at 300 K, measured for $I \parallel a$ and $I \parallel c$, respectively, at zero pressure is shown in Figure 1a. The data measured under pressure are shown in Figure 1b, at the selected pressures of 0.2, 1.0 and 1.8 GPa. Hydrostatic pressure results in rather opposite effects for $I \parallel a$ and $I \parallel c$. For $I \parallel a$, pressure leads to an overall reduction of $r_a(T)$ and a recovery of the FL $T^2$ term at low temperatures (see section IV), whereas for $I \parallel c$ $r_c(T)$ increases and develops a relative minimum at low temperatures ($T_{min} \sim$ 4.8 K at 1.8 GPa). The anisotropy in the resistivity of $U_2Pt_2In$ increases as a function of pressure. In Figure 2a, we show the low-temperature data taken in the interval 0.3-15 K for $I \parallel a$. For both current directions $r_0$ shows moderate changes as a function of pressure, which is another indication that the high $r_0$-values are not exclusively due to disorder.

The data shown in Figure 1b were taken on one and the same sample (#1), which had a platelet-like shape, such that the current could be applied along the $a$ and the $c$ axis. The zero-pressure data (Figure 1a) for $I \parallel a$ were also measured on this sample, while the data for $I \parallel c$ were measured on a second crystal (sample #2). Measurements under pressure on other single crystals with $I // c$ (sample #3) and with $I \parallel a$ (sample #4), confirm the overall behavior: an increase of the transport anisotropy, the development of a low-temperature minimum in $r_c(T)$ and the recovery of a $T^2$ term in $r_a(T)$. Although all crystals were cut from the same single-crystalline batch, there is a weak sample dependence of some of the resistivity features, especially the values of $T_{min}$ are different for samples #1 and #3. For sample #1, the minimum develops near 1.0 GPa and attains a value of $T_{min} \sim 4.8$ K at 1.8 GPa, while for sample #3, the minimum develops near 1.2 GPa and attains a value of $T_{min} \sim 2.1$ K at 1.8 GPa [21].

The normalized electrical resistivity of single-crystalline $U_2Pd_2In$ at pressures of 0.2, 1.0 and 1.8 GPa is shown in Figure 3 for $T < 50$ K. The data were obtained for a current direction along [101]. The overall shape of the resistivity curve does hardly change with pressure. However, $T_N$ decreases slightly. The inset shows the pressure variation of $T_N$ as measured by the temperature of the maximum in $r(T)$. $T_N$ shows a small decrease from 37.5±0.5 K at zero pressure to 35.2±0.5 K at $p = 1.8$ GPa.

In order to determine the compressibility, the lattice constants *a* and *c* were measured as a function of pressure up to 7 GPa. In this pressure range both the *a* and *c* parameters decrease linearly with pressure. The uniaxial compressibility along the *c* axis $k_c = 2.42 \times 10^{-3}$ GPa$^{-1}$ is slightly larger than the one along the a-axis $k_a = 2.20 \times 10^{-3}$ GPa$^{-1}$. The volume compressibility amounts to $k = 6.82 \times 10^{-3}$ GPa$^{-1}$ ($= 0.682$ Mbar$^{-1}$). Thus the *c/a* ratio decreases with pressure, albeit at a very small rate. In the pressure range relevant for the current experiments (1.8 GPa) it decreases only by 0.04%.

## IV. ANALYSIS

One of the main results of the pressure experiments is the recovery of the FL behavior at moderate pressures for $I \parallel a$. At low temperatures the resistivity for $I \parallel a$ can be expressed as $r = r_0 + aT^\alpha$. We have evaluated the exponent *a* by calculating

$$a = 1 + \frac{\mathrm{d}\ln\left(\frac{\mathrm{d}\rho}{\mathrm{d}T}\right)}{\mathrm{d}\ln T} \qquad (1)$$

which eliminates the uncertainty in the value of $r_0$. In fact, by computing *a* as a function of temperature with help of eq.(1) an effective $a_{eff}(T)$ is obtained. At the lowest temperatures $a_{eff}(T)$ attains a constant value. The resulting values of $a(p)$ for $T \rightarrow 0$ are shown in Figure 2b. At zero pressure $a = 1.25 \pm 0.05$, but under pressure *a* increases and attains a value of $2.0 \pm 0.1$ at $p \sim 1.0$ GPa. At still higher pressures the value of *a* remains constant, while the Fermi-liquid temperature $T_{FL}$ below which the $r \sim T^2$ is observed increases up to 1.5 K at the maximum pressure. The coefficient of the $T^2$ term amounts to $2.1 \pm 0.2$ μΩcm/K$^2$ at 1.0 GPa and decreases to a value of 0.40±0.04 µΩcm/K$^2$ at 1.8 GPa. From these coefficients we can obtain a rough estimate for the electronic specific-heat coefficient by using the Kadowaki-Woods relation [24]. The resulting γ-values are 0.46±0.02 J/molU-K$^2$ at 1.0 GPa and 0.20±0.01 J/molU-K$^2$ at 1.8 GPa. The large γ-value at 1.0 GPa is in-line with the heavy-fermion desciption of U$_2$Pt$_2$In and is of same order as the value of c/T at 1 K (0.41 J/molU-K$^2$ [2]) at ambient pressure.

According to the magnetotransport theory of Rosch [15], $T_{FL}$ is a function of the distance (measured by the pressure) to the QCP and varies initially as $T_{FL} = a_1 (p-p_c)$ with a cross-over to $T_{FL} = a_2 (p-p_c)^{1/2}$ at higher distances, where $p_c$ is the pressure at the QCP. The pressure ranges in which the different laws are observed depend on the amount of disorder $x$ in the system ($x \approx 1/RRR$). For $I \parallel a$, we measure $x \sim 0.6$, which indicates that our sample is in an intermediate regime of disorder. This implies that at the QCP the $\Delta r \sim t^{3/2}$ law (regime I) is strongly suppressed, and the low-temperature resistivity is dominated by a $\Delta r \sim tx^{1/2}$ law (regime II). However, under pressure the behavior $\Delta r \sim t^2 r^{-1/2}$ for $T \rightarrow 0$ (regime III) becomes more and more dominant. In Figure 4a we show regime II and III as deduced by fitting the resistivity under pressure to a $T^2$ term at the lowest temperatures and a term linear in $T$ at higher temperatures. An example (at 1.8 GPa) of the quality of such fits is shown in Figure 4b. Figure 4a shows that $T_{FL}$ is a linear function of pressure with $p_c = 0$, in agreement with the theory of Rosch. The cross-over to a $T_{FL} = a_2 (p-p_c)^{1/2}$ dependence (dashed line in Figure 4a) is expected near 3.0 GPa. The $r \sim T$ region is predicted to occur in the reduced temperature range $x < T/G < x^{1/2}$ ($x < 1$), where $G$ defines the temperature scale where the spin fluctuations are destroyed ($G \sim T_K$ or $T_{coh}$). From Figure 4a we extract that the $r \sim T$ region is found in the temperature range 2.8-4.7 K, from which it follows $x = 0.34$ and $G = 8.1$ K. The agreement between the calculated value $x = 0.34$ and the experimental value $x \sim 0.6$ (=1/RRR) is, given the rather simple data treatment, satisfactory. Notice that $G$ is about equal to $T_m$, i.e. the temperature of the maximum in the susceptibility. The distance to the QCP is given by $r = \zeta p$ with $\zeta = 0.11$ GPa$^{-1}$. We conclude that the temperature-pressure diagram presented in Figure 4a is in good agreement with the scaling diagram for the resistivity presented by Rosch, except for the value of the exponent $a = 3/2$ predicted for region I. Instead we find $a = 1.25$ at $p = 0$ and $1 < a < 2$ for non-zero pressures. However, under pressure the $\Delta r \sim t^{3/2}$ region becomes very small and its proper observation is hampered by the cross-over from regime III to II. Therefore, the resistivity data under pressure for $I \parallel a$ are consistent with $U_2Pt_2In$ exhibiting an antiferromagnetic QCP at zero pressure.

For the electrical resistivity measured for $I \parallel c$ the situation is different. At zero-pressure $r_c \sim T^\alpha$ with $\alpha \sim 0.9$ for $T \to 0$. Under pressure $\alpha$ first increases, but near 1.0 GPa $r_c(T)$ develops a minimum, which becomes more pronounced with increasing pressure. This behavior, first observed on sample #3, was reported in Ref.[21]. Measurements of $r_c(T)$ at $p = 1.8$ GPa in a magnetic field applied along the current direction show a suppression of $T_{min}$ from ~ 4.8 K at zero field to ~ 2.2 K in a field of 8 T, which confirms the magnetic nature of the minimum (see also Ref.[21] for data on sample #3). For $I \parallel c$, $x = 0.9$ (RRR = 1.1) and the sample is in the regime of strong disorder. As a consequence, the regime $r \sim T$ should be suppressed. Indeed we find $\alpha < 1$. For $T \to 0$ $\alpha \sim 0.9$, while at finite temperatures $\alpha_{eff}(T)$ is even smaller. Clearly, for $I \parallel c$ disorder is large and the magnetotransport theory of Rosch does not apply.

An anisotropic resistivity is also evident at higher temperatures. The resistivity shows a weak maximum at $T_{max}$~70 K for $I \parallel a$ and ~100 K for $I \parallel c$. The pressure effect on $T_{max}$ is strongly current-direction dependent: $T_{max}$ shows a strong increase for $I \parallel a$, while it decreases slightly for $I \parallel c$. For $I \parallel a$ we suggest that the maximum is due to the formation of the Kondo-lattice, in which case $T_{max}$ is proportional to the Kondo temperature $T_K$ [25]. Neglecting the (weak) phonon contribution to the resistivity, the pressure data presented in Figure 1b can be used to calculate an empirical Grüneisen parameter for $T_K$ defined as

$$G_K = -\left.\frac{\partial \ln T_K}{\partial \ln V}\right|_{V=V_0} \tag{2}$$

where $V_0 = V(p=0)$. Since $T_{max} \propto T_K$, it follows [26]

$$G_K = -\left(\frac{V-V_0}{V_0}\right)^{-1} \ln \frac{T_{max}(p)}{T_{max}(0)} \tag{3}$$

The relative volume change $(V-V_0)/V_0$ is given by $\Delta V/V_0 = -\kappa p$. For $I \parallel a$ we estimate $G_{K,a} = 49.7 \pm 7.1$. This value of $G_K$ is close to the values $G_K = 59$ and 65 [26], reported for other Kondo-lattice systems, like $CeInCu_2$ and $CeCu_6$, respectively. This indicates that the strong pressure dependence of $T_K$ as deduced from the electrical resistivity of $U_2Pt_2In$ measured for $I \parallel a$ is not unusual. The slight decrease of $T_{max}$ with pressure for $I \parallel c$, on the other hand, is unexpected. A similar analysis of the pressure dependence of $T_{max}$ as performed for $I \parallel a$, results in $G_{K,c}$ = -6.8±0.7. Such a large anisotropy of $G_K$ is unlikely. Therefore, we suggest that $r_c(T)$ at high temperatures is not dominated by the Kondo effect, but by another mechanism, which results in an enhancement of $r_c(T)$ under pressure. On the other hand the enhancement of $r_c(T)$ under pressure might be related to a very strong Fermi surface anisotropy.

## V. DISCUSSION

The pressure dependence of the electrical resistivity of $U_2Pt_2In$ for $I \parallel a$ as measured up to 1.8 GPa, can satisfactorily be analyzed within the scenario of an antiferromagnetic quantum phase transition at $T = 0$. At low temperatures, pressure results in the recovery of the FL regime in agreement with the transport theory of disordered metals near an antiferromagnetic QCP model proposed by Rosch, while at high temperatures the pressure-induced shift of $T_{max}$ can be attributed to the usual increase of the Kondo temperature. The increase of $T_K$ reflects a stronger conduction electron - $f$-electron hybridization and, therefore, the exchange parameter $J$ increases. This is in agreement with the appearance of a FL $r \sim T^2$ behavior at low temperatures.

An estimate of the relative increase of $J$ under pressure can be deduced by calculating the hybridization matrix element for the total conduction electron hybridization at the $f$ atom, $V_{cf} = (V_{df}^2 + V_{pf}^2 + V_{ff}^2)^{1/2}$ [27,28], with $J \propto V_{cf}^2/(E_F - E_f)$ as in the Coqblin-Schrieffer model [29]. In a first simple approximation the distance $E_F$-$E_f$, which measures the energy distance of the $f$ level ($E_f$) relative to the Fermi energy ($E_F$), can be taken as a constant, since the $f$ level is stable with respect to the Fermi energy. Since the compressibility is nearly isotropic, and assuming that the co-ordinates of the atoms in the unit cell do not change with pressure, the primary effect of pressure is a uniform reduction of the inter-atomic distances. With the compressibility value $k$ = 6.82x10$^{-3}$ GPa$^{-1}$, we calculate an increase of $V_{cf}$ of 2.3% in the pressure range 0-1.8 GPa. Thus by applying a moderate pressure of 1.8 GPa, $U_2Pt_2In$ is shifted considerably into the non-magnetic region of the Doniach-like diagram. A similar calculation for the antiferromagnet $U_2Pd_2In$ results in an increase of $V_{cf}$ of 2.4% at 1.8 GPa, where we have used the same compressibility value as for $U_2Pt_2In$ (the compressibility of $U_2Pd_2In$ has not been measured so far). Notice that the difference in the crystallographic structures of $U_2Pt_2In$ ($Zr_3Al_2$-type) and $U_2Pd_2In$ ($U_3Si_2$-type) can be neglected in the calculation of $V_{cf}$ [7]. In Figure 5 we show a tentative Doniach-like diagram for the compounds $U_2Pt_2In$ and $U_2Pd_2In$ under pressure. For $U_2Pt_2In$ we show $T_{FL}$, while for $U_2Pd_2In$ we show $T_N$ as a function of $V_{cf}^2$. The weak variation of $T_N$ of $U_2Pd_2In$ under pressure is consistent with its location near the maximum ordering temperature in the Doniach phase diagram for the $U_2T_2In$ family of compounds. According to Figure 5, a very rough estimate of the pressure needed to tune $U_2Pd_2In$ to a QCP is ~ 7 GPa.

An attractive method to probe the Doniach-like diagram further is by expanding the lattice of $U_2Pt_2In$ through alloying with e.g. Th, which should result in magnetic order. From the change of the lattice constants in the pseudoternary series $(U_{1-x}Th_x)_2Pt_2In$ [30], we calculate a negative chemical pressure of -0.2 GPa per at.% Th doping. Thus for $x = 0.1$, the negative chemical pressure amounts to -2 GPa, which should lead to an ordering temperature in the range of 15-20 K based on Figure 5. Resistivity studies on polycrystalline $(U_{1-x}Th_x)_2Pt_2In$ samples reported in the literature [31], indicate that the resistivity at low temperatures ($T > 1.5$ K) gradually changes from $\mathbf{r} \sim T$ towards $\mathbf{r} \sim T^2$ as the Th content increases. For $x = 0.1$ the resistivity data show a change of slope near 19 K, which possibly indicates magnetic ordering. However, the change of slope might also be due to small amounts of impurity phases, like UPt, which has two magnetic phase transitions at 27 K and 19 K [32]. Specific-heat measurements carried out on polycrystalline $(U_{1-x}Th_x)_2Pt_2In$ ($0 \leq x \leq 0.1$) [33] do not show any evidence for magnetic order down to 2 K. On the other hand, µSR experiments do signal magnetic transitions in some of the samples [7]. These conflicting results evoke the need for high-quality single-phase material. Also, it should be noticed that substitution of U by Th dilutes the *f*-electron lattice, which might impede the emergence of magnetic order.

Although the analysis of the resistivity ($I \parallel a$) of $U_2Pt_2In$ under pressure is consistent with an antiferromagnetic QCP in 3D, the divergency of the specific heat, $c/T \sim -\ln(T/T_0)$ rather indicates a ferromagnetic QCP. However, a diverging $T\ln T$ term in the specific heat is a general feature of a system with a dimension *d* equal to the dynamical critical exponent *z*. Possibly, quasi-two dimensional fluctuations could lead to a reduction of *d* and *z*, like for the NFL compound $CeCu_{5.9}Au_{0.1}$, which is located at an antiferromagnetic QCP and for which it has been proposed $d \approx z \approx 2.5$ [6]. Inelastic-neutron scattering experiments could shed light on this issue.

The rapid recovery of the FL behavior under pressure as probed by the resistivity data for $I \parallel a$ does not yield support for Kondo-disorder as mechanism for NFL behavior in $U_2Pt_2In$. Since the compressibility is isotropic, pressure is expected to result in the further broadening of the distribution of Kondo-temperatures and thus the concurrent NFL behavior is preserved.

The strong current-direction dependence of the pressure effect is unusual. At zero pressure the data indicate a significant anisotropy of the Fermi surface. Under pressure this anisotropy becomes even stronger. The emergence of a minimum in the resistivity for $I \parallel c$ is not understood. By comparing the unit cell volumes and the $c/a$ ratios of $U_2Pt_2In$, $U_2Pd_2In$ ($T_N = 37.5$ K) and $U_2Pd_2Sn$ ($T_N = 15$ K) [1], we proposed [21] that $T_{min}$ could indicate magnetic ordering of the spin-density wave type, if the $c/a$ ratio varies strongly with pressure and acts as control parameter in the Doniach diagram, rather than the unit cell volume. However, our new compressibility data show that $c/a$ hardly changes with pressure, which invalidates this hypothesis.

## VI. CONCLUSIONS

We have investigated the effect of hydrostatic pressure ($p \leq 1.8$ GPa) on the non-Fermi liquid state of $U_2Pt_2In$, by means of electrical resistivity experiments in the temperature interval 0.3-300 K. The experiments carried out on single-crystals show that the pressure effect depends strongly on the current direction. For $I \parallel a$, the low-temperature resistivity at zero pressure shows a NFL power law behavior, $r \sim T^{\alpha}$, with $a = 1.25 \pm 0.05$. Under pressure the NFL behavior is suppressed: $a$ increases and attains the FL value of $2.0 \pm 0.1$ at $p \sim 1.0$ GPa. The data for $I \parallel a$ can be analyzed satisfactorily within the magnetotransport theory of Rosch, which provides strong evidence for the location of $U_2Pt_2In$ at an antiferromagnetic quantum critical point. From the pressure-induced shift of the high-temperature maximum in the resistivity, we conclude that $T_K$ varies strongly with the volume ($G_K \sim 50$), which is consistent with the rapid recovery of the FL term under pressure at low temperatures. For $I \parallel c$ the behavior is complex and the data suggest the enhancement under pressure of an additional component to the resistivity. We have also measured the effect of pressure on the antiferromagnetic ordering temperature ($T_N = 37$ K) of the related compound $U_2Pd_2In$. The variation of $T_{FL}$ of $U_2Pt_2In$ and $T_N$ of $U_2Pd_2In$ under pressure can be described in a simple Doniach-type phase diagram.

## ACKNOWLEDGMENTS

P.E. acknowledges the European Commission for a Marie Curie Fellowship within the TMR program. T.N. thanks A. Matsushita for assistance in developing the high-pressure cell. The authors acknowledge K. Prokeš for providing the crystal of $U_2Pd_2In$.

**FIGURE CAPTIONS**

Fig. 1    Temperature dependence of the normalized resistivity of $U_2Pt_2In$ for $I \parallel a$ and $I \parallel c$: (a) at zero pressure and (b) under pressures as indicated. Notice the $\log T$ scale.

Fig. 2    (a) Low-temperature dependence of the normalized resistivity of $U_2Pt_2In$ for $I \parallel a$ at different pressures.

(b) Pressure dependence of the resistivity exponent $a$ ($r \sim T^{\alpha}$) for $I \parallel a$. The solid line is to guide the eye.

Fig. 3    Temperature dependence of the normalized resistivity of $U_2Pd_2In$ at different pressures as indicated. Insert: Pressure dependence of $T_N$. The line is a guide to the eye.

Fig. 4    (a) Pressure dependence of $T_{FL}$ (¦) and of the temperature range in which $r \sim T$ (between (o) and (x)). The solid lines delineate regimes I, II and III (see text). The meaning of the dashed line is explained in the text.

(b) Temperature variation of $\Delta r/r_{RT}$ for $U_2Pt_2In$ at $p = 1.8$ GPa. The dashed and solid lines show the behavior $r \sim T^2$ and $r \sim T$.

Fig. 5    Doniach-type diagram for $U_2Pt_2In$ and $U_2Pd_2In$ at zero pressure (closed symbols) and under pressure (open symbols). AF= antiferromagnetic order, FL = Fermi-liquid regime. The lines serve to guide the eye.

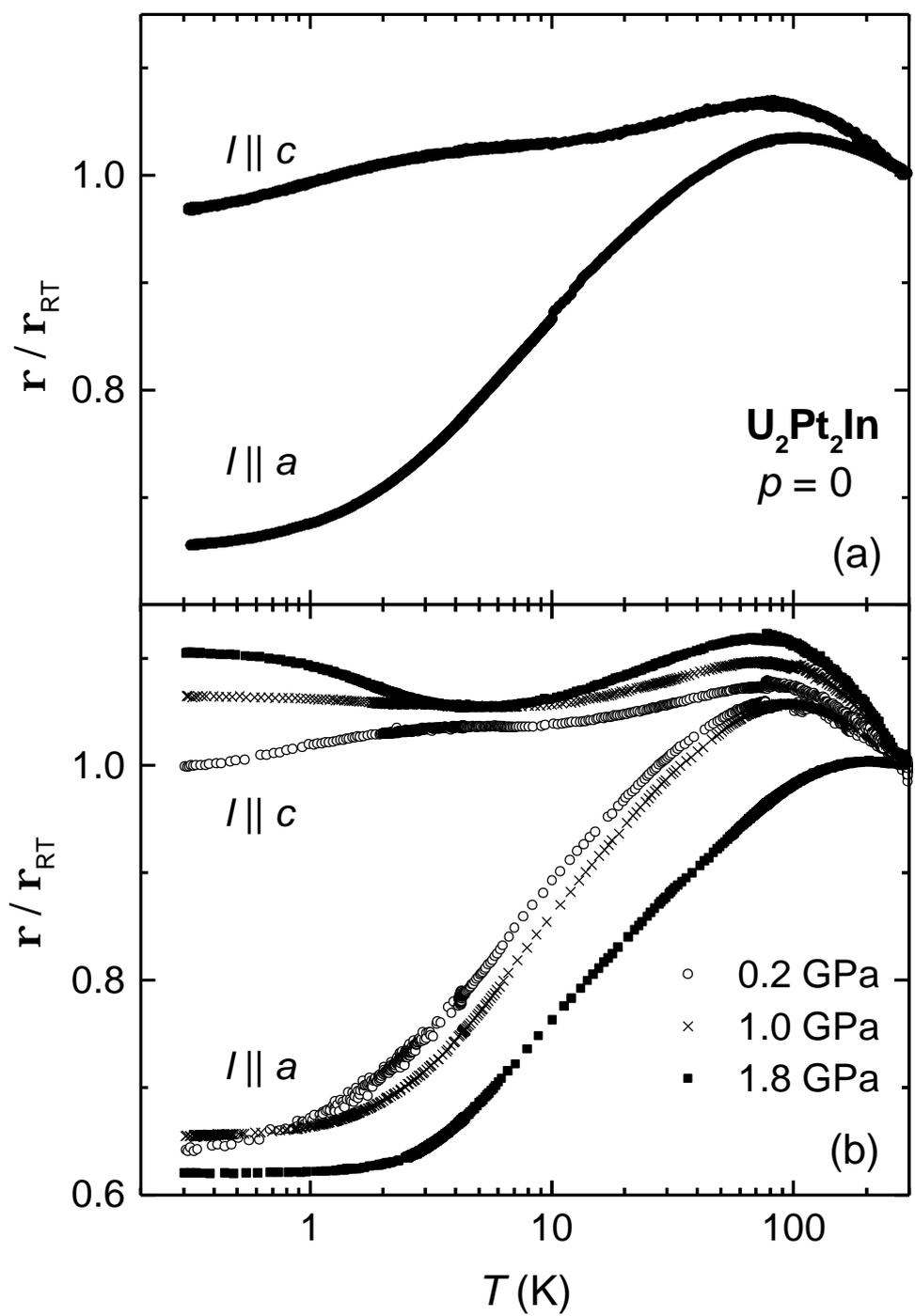

**Figure 1**

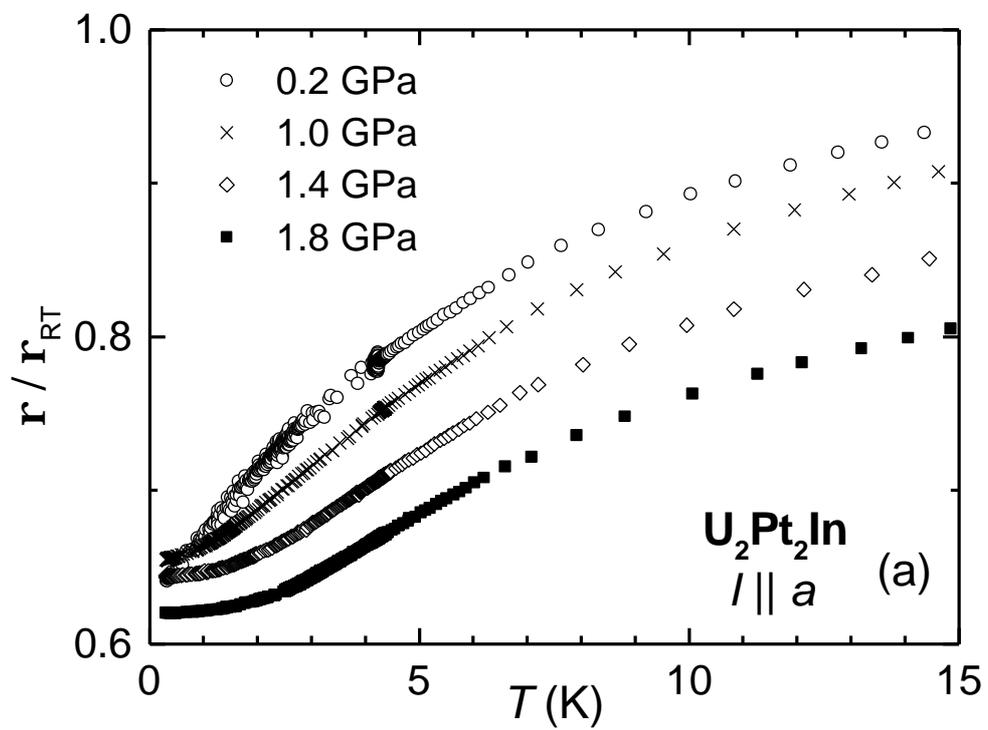

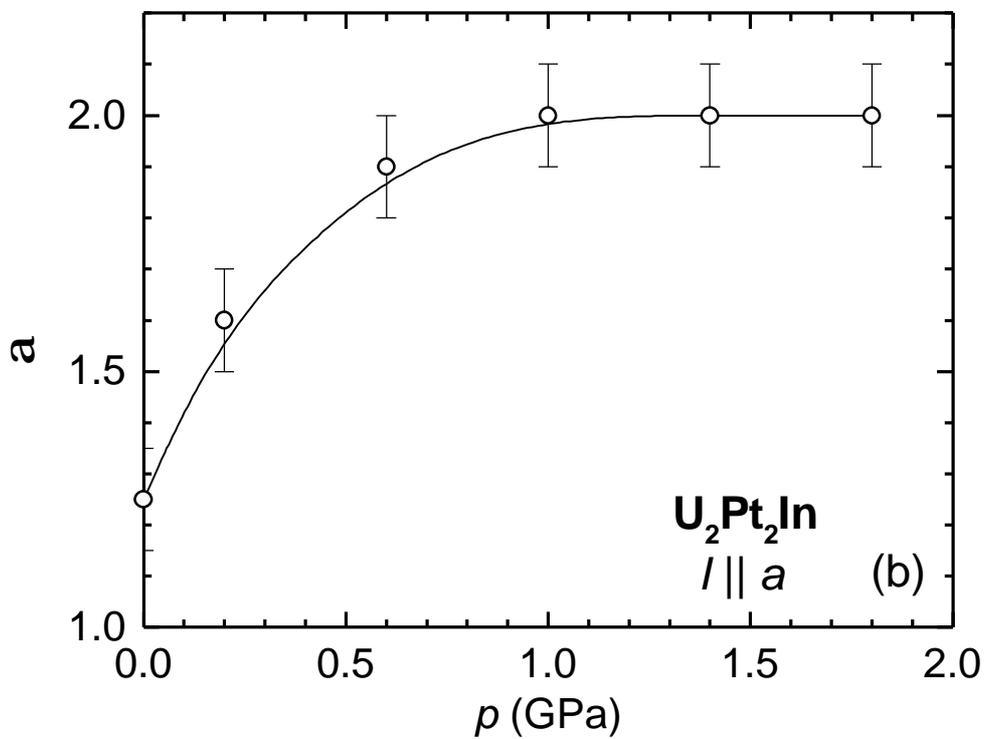

**Figure 2**

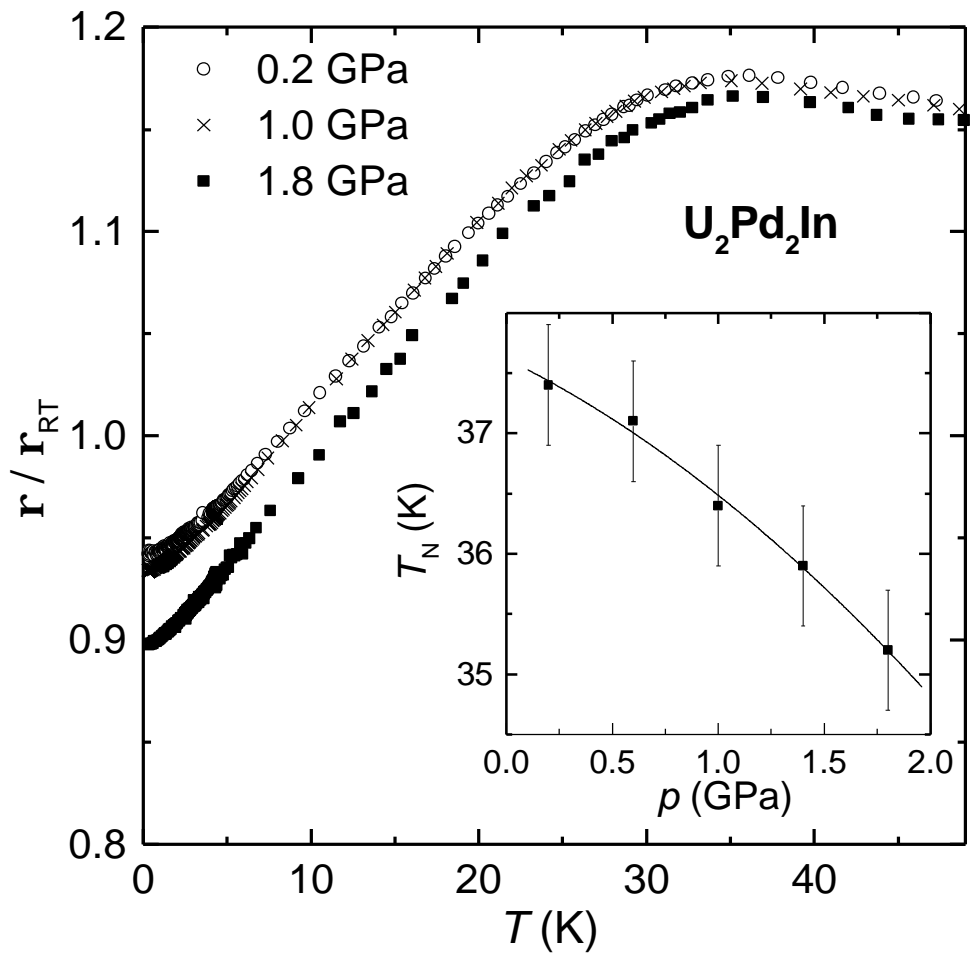

**Figure 3**

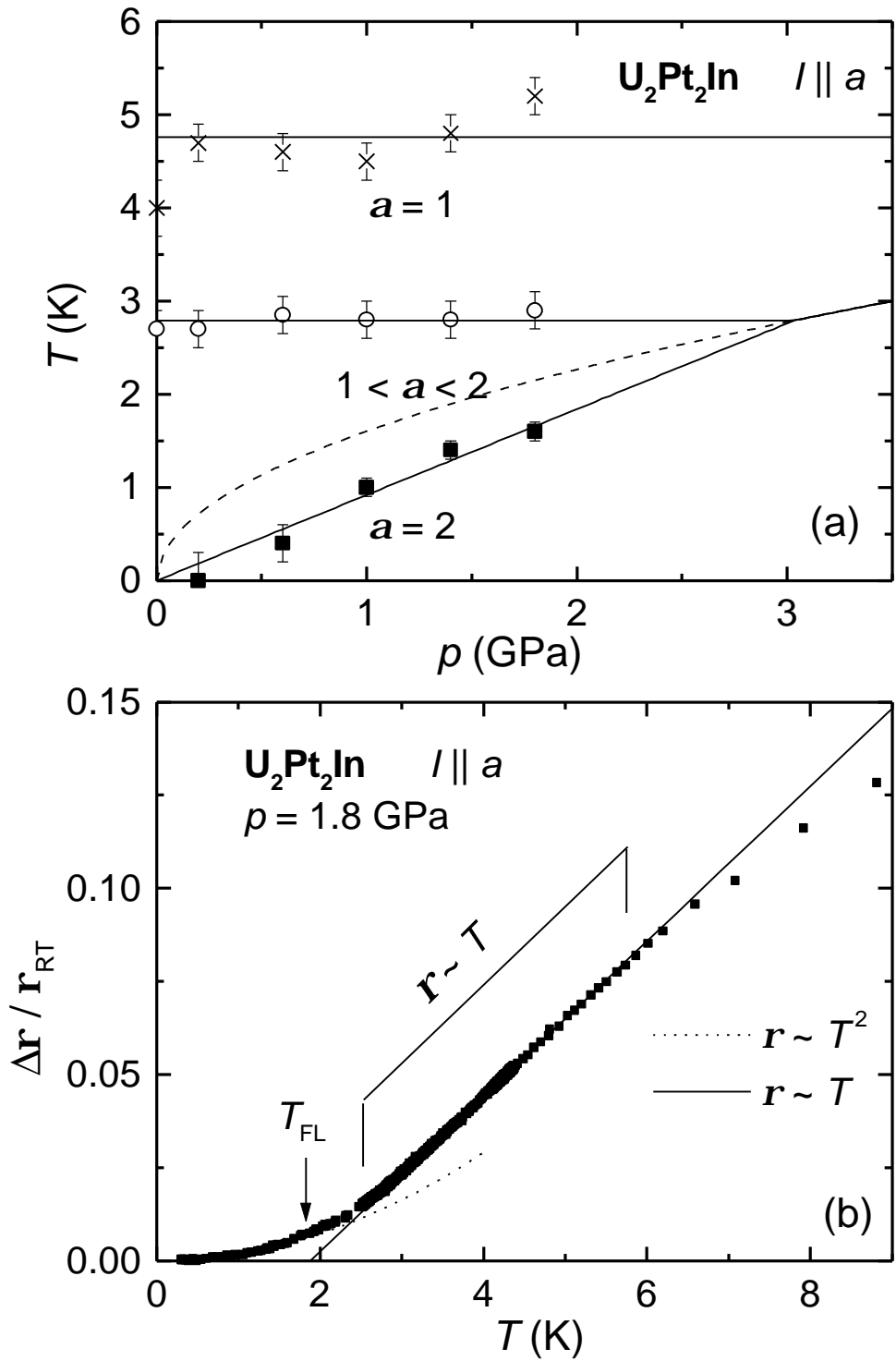

**Figure 4**

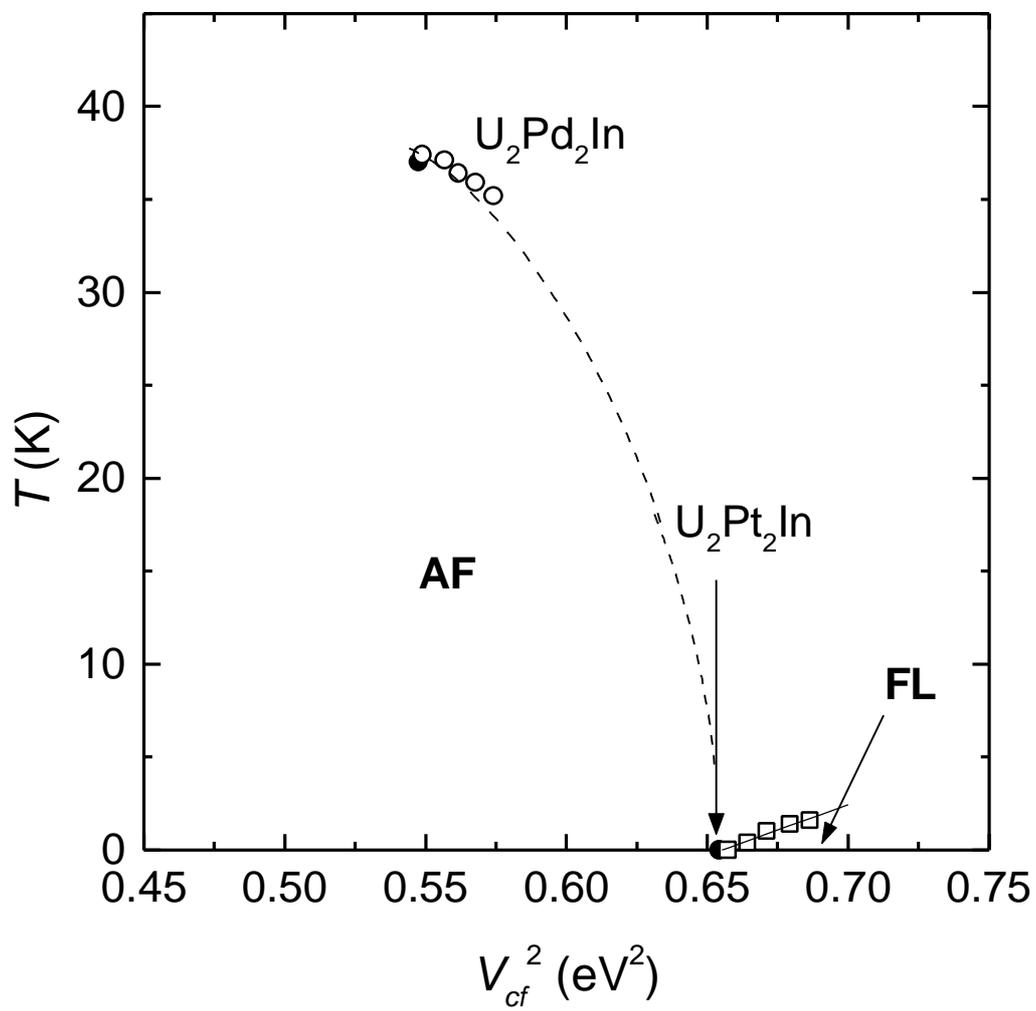

**Figure 5**